\documentclass{article}
\usepackage[utf8]{inputenc}
\usepackage{amsmath}
\usepackage{graphicx}
\usepackage{complexity}
\usepackage{hyperref}

\title{Is Complexity Important for Philosophy of Mind?\footnote{PREPRINT VERSION.}}
\author{Kristina \v{S}ekrst\footnote{University of Zagreb, ksekrst@ffzg.hr}  \and Sandro Skansi\footnote{University of Zagreb, sskansi@hrstud.hr}}
\date{October 2021}

\newtheorem{thesis}{T.}

\begin{document}

\maketitle

\section{Introduction}
In this chapter, we explore the intricacies of computational complexity and its applications to philosophy of mind. Our starting point is the Church-Turing thesis, which in its most-quoted form \cite{church, turing-thesis} says:

\begin{thesis}
Anything computable is computable by a Turing machine.
\end{thesis}

Even though it is only a thesis, not just unproven, but by definition, unprovable, the Church-Turing thesis has spearheaded research in computability for decades. At its core, it is a definition, whose task is to define in precise terms what it means to be ``computable''. But it is not called a ``definition'' because it is much more than just a definition, begging to be falsified by a counterexample, much like a conjecture. It could be argued that the Church-Turing thesis is a challenge set forth to redefine computation should anything noteworthy be discovered. But nothing was. In almost a century now, many new models of computation have arisen, and all were eventually proven to be equivalent to the Turing machine.

Although there were many more or less simplistic attempts to relate Turing machines to minds, we will argue that the simpler arguments along those lines do little to clarify the concept of mind. As for more complex arguments, they cannot be provided right away, for a rough framework along the lines of Ashby-style cybernetics \cite{ashby} is to be acquired in order to put forth any sensical, i.e. non-dualist, comparison between minds and machines.

Once this is done, the notions of computational complexity, and its generalization -- metaphysical complexity -- can be rediscovered in the mind, and, surprisingly, in society. Even though computation-in-society is traditionally viewed as a strong cybernetic moment, it is not that unique to cybernetics and has become commonplace in mainstream AI as well (cf. \cite{minsky}).

One could argue that this, quintessentially non-dualist position has strange consequences, and in fact, one would be right. But we consider this not as grounds for dismissal, but an intriguing new theory that has a certain level of compactness and soundness, and the unusual consequences it entails are solid and should spark interest, not dismissal.

\section{Church-Turing thesis revisited}

Even though cybernetics today has lost much of its former shine, there is a certain ``hipster'' vibe to it.\footnote{ A reader interested in philosophical cybernetics should see \cite{ashby}, \cite{skansi-sekrst}.} A fully fledged cybernetic approach is well beyond the scope of this chapter, but we will need an element of it to modify the Church-Turing thesis, which will reopen the subject.

The main problem we have with directly applying the Church-Turing thesis to form a theory of mind is its essentially dualist language: it speaks of computation, whereas the mind does not compute. But, the mind does something more general than computing: it solves problems, and computers also solve problems. And as nontrivial computations are achieved by nontrivial program code, so are nontrivial solutions to problems achieved by inventing, redefining or even just following complex procedures. We should note that this generalization wholly subsumes the original Church-Turing thesis. By rephrasing the Church-Turing thesis in terms of \textit{solvability} we get:

\begin{thesis}
Anything solvable is solvable by a Turing machine.
\end{thesis}

Even though it looks deceivably simple, and as a trivial rephrasing of the original thesis, this modification goes a long way, as we will try to show in the rest of the paper.

One might raise an objection here. The original thesis was strong due to the fact that \textit{computability} is synonymous with \textit{computation}, and it comes to no great surprise that anything that computes can, in fact, be rephrased as a Turing machine. But solvability is not something that is ``mechanical'', and the broadening of the scope should not be taken lightly, as there might be problems that are solvable, but not with a Turing machine. Take, for example, consoling someone who is sad. In a sense, it could be said that one could solve a problem just by doing ``her magic'', without an overt attention to some procedure.

To see why this argumentation is flawed, notice that solving a problem by accident is not a solution to the problem. More precisely: even though a problem might be \textit{solved} by accident, we would not accept the idea that this constitutes a good way to go about and solve similar problems. One's financial problems might be solved by winning the lottery, and yet we do not consider playing the lottery as a valid alternative to financial responsibility.

But a stronger point could be made: this kind of an argument does not arise from our modification of the Church-Turing thesis, for it could have been made against the original thesis as well. Consider a program that returns random numbers from 0 to 18, when given two-single digit inputs (ranging from 0 to 9). This program occasionally ``computes'' addition for single digit numbers, but we do not consider this to be ``computation'' because it does not have a procedure for addition. Whether we should think like this is irrelevant for our point: the only thing that is important is that our generalization of the Church-Turing thesis, which will enable us to apply computational complexity to minds, does not have \textit{prima facie} leakages. But before we can turn our attention to this application, we must revisit complexity to get a glimpse of what we can get from it.

\section{Complexity}

We ask the reader to multiply 12 by 2. Now we ask you to multiply 1245 by 12. You'd be expected to perform significantly slower when presented with such a task. If you posit the same problem to a computer, you would expect it to be fast. But what if we increased our bets, and tried to find all prime numbers less than 10? That's not a big challenge for a man, let alone a computer. Now, please, present all prime numbers less than a billion. That's some work even for a computer. Of course, it would still be measured in seconds or minutes, but the first problem would be solved instantly.

Intuitively, it is easy to check whether a certain thousand-digit number is prime, but it is not easy to calculate such a number. That is where computational complexity comes into play. Computational complexity theory is concerned with how much computational resources are required to solve a given task \cite{arora-barak}, i.e. it analyzes and classifies computational problems according to their \textit{inherent difficulty}.

In computer science, \textit{computational complexity} refers to the amount of resources – either temporal or spatial – required to run an algorithm. As we have mentioned, computational complexity theory studies and analyzes algorithms and problems. Typically, the computational efficiency is measured as a number of basic operations an algorithm performs as a function of its input length \cite{arora-barak}. The complexity of a certain problem is an upper bound, i.e. a function $n \mapsto f(n)$, where $n$ is the size of the input, and $f(n)$ is the worst-case complexity, the maximum amount of resources needed for the computation to finish. A problem is regarded as inherently difficult if it requires significant temporal (long running time) or spatial (great memory requests) resources.  

Suppose that you are organizing housing accommodations for a group of 400 philosophers invited to a conference regarding philosophy of mind. Since space is limited, only 100 lucky ones will receive their places, and since some dualists do not get along with some physicalists, you have also received a list of incompatible pairs. Our goal is to calculate $\binom{400}{100} = \frac{400!}{(400 - 100)! \times 100!} = 3 \times 10^{27}$. That means the possible number of combinations is around $3 \times 10^{27}$. To compare, as of June 2020, the fastest supercomputer Fugaku achieved a score of 415 PFLOPS \cite{vegh}, meaning 100 peta FLOPS. A 1 PFLOP system is capable of performing one quadrillion ($10^{15}$) floating-point operations per second. Even if we needed just one operation (and in practice, it is much more) to check each possible combination, the difference between $10^{18}$ operations per second and $10^{27}$ possible combinations is still $10^9$ seconds. It does not seem bad at first glance, but that is more than 30 years. If we used a supercomputer with a processing power of ``just'' $10^{15}$ floating-point operations per second, it would take us more than 30 000 years to try out all the possible combinations. In this case, it is not the philosophers’ fault. It seems to be inherent to properties of space and time themselves. Just because a function is computable in the Church-Turing sense does not automatically mean it is computable in the real world, and it does not automatically mean it actually solves the problem at hand. Remember that example in which a second-grade supercomputer would need 30 000 years to come up with a solution to our problem. Yes, it is \textit{computable}, but at what cost?  

\begin{figure}
    \centering
    \includegraphics[width=\textwidth]{cc}
    \caption{Big-O notation is used to classify problems taking into account how time/space requirements grow with input size. \newline \small Source: Wikimedia Commons, Creative Commons Attribution-Share Alike 4.0 International.\normalsize}
    \label{cc}
\end{figure}

\textit{Time complexity} describes an amount of time it takes to run an algorithm. In the case of \textit{constant time} -- O($1$) -- the upper bound for the running time of such a calculation does not depend on the size of an input. For example, if we had to point to a first member of the list, it does not matter whether the list has 5 or 5 billion members. \textit{Logarithmic time} -- O($\log{n}$) -- is considered highly efficient since the ratio of operations needed decreases compared to the size of the input $n$ and tends to zero. Examples in computer science include various search algorithms such as binary search trees. \textit{Linear time} -- O($n$) -- means that the running time increases at most linearly with the size of the input $n$, for example, finding an element within a list by checking each element until the match is found. \textit{Polynomial time} -- O($n^2$) -- is upper bound by some polynomial expression in the size of the input $n$, and includes many natural problems such as various ways of sorting a list. \textit{Exponential time} -- O($2^{n^k}$) -- would grow significantly faster than any polynomial function. Exponential functions are found everywhere around us, from the famous wheat and chessboard problem to modern day inflation, compound rates, and recently in the case of some COVID infection rates.

Back to our housing problem. A mentioned list with all the possible combinations would be easy to verify, but, as we have seen, it is not easy to acquire such an answer, and time complexity would grow to unusable limits. This is an example of an $\NP$ problem (\textit{nondeterministic polynomial time}), which is a category of decision problems.  A \textit{decision problem} is a problem with a yes/no answer. Is $n$ a prime number? Yes. Is $m$ a prime number? No. In computational complexity theory, $\NP$ comprises decision problems which are \textit{verifiable} in polynomial time by a deterministic Turing machine, while class $\P$  consists of problems that are \textit{solvable} deterministically in polynomial time. $\NP$-hard problems are as hard as $\NP$ problems, but they do not have to be decision problems. For example, instead of asking is there (yes/no) a shortest route between points that is reachable by visiting each point only once, one could just ask to find the shortest possible route.

$\NP$-complete problems are $\NP$ problems that could be (eventually) solved by brute-force algorithms, whose solution can be \textit{verified} quickly, and the solution to any $\NP$-complete problem could be used \textit{to simulate any other problem of the same class}. It is interesting to observe that the first problem proven to be an $\NP$-complete problem \cite{cook, levin} was in fact a philosophical problem, i.e. that of logic. The \textit{Boolean satisfiability problem} ($\SAT$) is a decision problem of determining whether there exists an interpretation that satisfies a given Boolean formula.

We have explained the decision part and the polynomial part of the $\P$ and $\NP$ class definitions. This leaves us with \textit{nondeterministic}. One could inquire why we need nondeterminism in the mind? A possible answer was provided by Ashby \cite{ashby}, who states that it is a heuristic method that sustains variety. Nondeterministic Turing machines can also be seen as modifications of Turing machines to capture $\NP$, i.e. arising from natural $\NP$ problems, and not the other way around, as it is usually taught. This brings us to the biggest question in computer science today – the $\P$ vs. $\NP$ question: is the $\NP$ class reducible to a $\P$ class of problems, i.e. can every problem whose solution can be quickly \textit{verified} also be solved \textit{quickly}? In that case, if one would find a polynomial solution to an $\NP$-complete problem, it could be used as a solution for all of the problems in the same class, the same way cracking a safe leads you to the ability to crack all the safes that are the same model.

\section{Ignoring complexity: the bread and butter of philosophy of mind}

As Urquhart \cite{urquhart} and Aaronson \cite{aaronson2011} rightly emphasize, philosophical treatments of the concept of computation often ignore issues related to complexity. Urquhart pinpoints that both Searle \cite{searle} and Churchlands \cite{churchlands} ignore questions of complexity and efficiency in their thought experiments since without considering the question of computational complexity, it is difficult to see how to determine functional brains at all. By taking computational resources into account, one can distinguish between objects that exist in the purely mathematical sense and devices that are physically constructible \cite{urquhart}. 

Intuitively, it seems that $\P$  problems are easy problems, and problems not in $\P$  (e.g. problems in $\NP$ and even more complex computational classes) are difficult. Such a postulate is actually \textit{Cobham-Edmonds thesis} \cite{cobham, edmonds} that asserts that computational problems are feasibly computed only if they lie in $\P$  i.e. if they can be computed in polynomial time.\footnote{Cobham-Edmonds's thesis restricts Church-Turing thesis. The latter states that solvable problems are invariant regarding different models of computations, but the former states that the time complexity of solvable problems is polynomially related.} $\P$  class implies easy and fast problems, while problems outside the $\P$ class (upwards in the computational hierarchy) imply difficult and slow computations.

One could ask whether complexity is only a matter of time, and in fact it is not. Complexity can be studied for any limited resource needed and it turns out, echoing Kant \cite{kant}, that the most fundamental are time and space. All other computational/metaphysical resources can be obtained with time and space, and time and space themselves are computationally connected, but, for now, it is beyond our knowledge whether time subsumes space. What we do know is that space subsumes time, and if the converse would turn out to be true, it would entail a space-time equivalence. We should note that there is a major metaphysical issue at stake here about true time and space and not just ``computational time and space'', which we need for our argument. One might say that this is a second cybernetic moment we have, but we beg to differ, since as with the first, it is a simple and intuitive insight, and if it turns out to be important for cybernetics as well, so much the better.  
Computational time and space are a subset of real time and space, but metaphysically speaking, they are fully representative of the real time and space, in the sense that all metaphysical properties of time and space are already present in computational time and space, and therefore studying computational time and space allows us to wholly study real time and space. We can also pose a challenge to the reader, and that is to find an essential property of real time or space that is not present in computational time and space, and therefore find a counterexample to our proposed assumption. We believe this to be an impossible task, in great part thanks to the Church-Turing thesis, but we welcome the opportunity to be proven wrong. 

The intuition whether computational time is metaphysically relevant might be strengthened with another computational theorem. \textit{Savitch's theorem} \cite{savitch} states that if a nondeterministic Turing machine can solve a problem using $f(n)$ space, then a deterministic Turing machine can do the same but in space squared: $\NSPACE(f(n)) \subseteq \DSPACE (f(n)^2$. That is, a deterministic Turing machine can simulate a non-deterministic Turing machine without needing much more space (notice it’s only quadratic, it does not rise exponentially!). Savitch's theorem is based on an algorithm for STCON or $st$-connectivity, a decision problem checking for vertices $s$ and $t$ in a directed graph, whether $t$ is reachable from $s$, i.e. determine whether there is a path between two vertices. Talking in Turing-machine terms, the algorithm goes through all the possibilities and checks whether the final tape (at the end of the execution) is reachable from the initial tape (when the program is started). The algorithm is based on recursion, but the key notion here is that space is \textit{reusable} and two recursive reachable computations are \textit{using the same memory}. Even though the time for this algorithm is very high, space is not. So, Savitch's theorem states that nondeterminism does not, in fact, help space-bounded computation.

We would like to pinpoint the metaphysical importance of \textit{reusability} here. Even though you can reuse space, you cannot reuse time. In the previous case, space does not grow exponentially, but computation itself might take much more time. It might be that in the case of time and space in computational complexity, time is more relevant for the question of presupposing an intelligent mind. If a person receives an answer in the Turing test (determining whether a computer is capable of acting or thinking like a human being \cite{turing}) by supercomputer cloud computation disseminated over thousands of computers taking a vast memory amount, but that answer is a fast one, we would be more tempted to postulate intelligence than in the case of a single brilliant computation with limited space, but with an increased amount of time.

By Savitch’s theorem, determinism and indeterminism are irrelevant to computational complexity when it comes down to space. There is at most an exponential difference between deterministic and nondeterministic Turing machines regarding their runtime, but there is at most a polynomial difference between their running space/size. Of course, if we have more space, we can compute more, but if time is bounded, space is bounded as well since we cannot access the space at hand. Imagine that you own a large factory. If you have, say, 10 years, you could manufacture (by yourself) a lot of items. But if you only have one hour, you could not even visit each workstation in the factory, let alone use it. But despite all the well-built intuition behind the reusability of space, we still do not know whether space helps more than time itself, that is, in computational complexity terms, whether $\P \stackrel{?}{=} \PSPACE$, where $\PSPACE$ is a set of all decision problems solvable by a Turing machine using a polynomial amount of space.

There is one more counterintuitive result. The \textit{Immerman–Szelepcsényi theorem} \cite{immerman, szelepcsenyi} states that nondeterministic spatial complexity classes are closed under complementation: $\NSPACE(s(n))$ = co-$\NSPACE(s(n))$ for any function $s(n) \geq \log{n}$. To explain, if a nondeterministic machine can solve a problem, another machine with the same resource bounds can solve its complementary problem (reversing the yes/no answer to a decision problem). In other words, proving that something does not exist (as a result of a computation, not in an ontological essentialist sense) is as easy as proving it exists. 

We have already noted the metaphysical importance of these ideas, but why are they important for the mind? By making the transition from computational time and space, we have distanced ourselves from something that purely affects computers to the more general ``real'' time and space, which affect both computers, minds, societies, nature, etc. In this way, we have extended the exact calculations of computational time and space and made them applicable (in theory) to anything which \textit{solves} anything in real time and space. The focus is now on procedures, solvability, and real time and space, but the instruments are those of computational complexity. The human mind is most certainly one such ``machine'', which has procedures, solves things, and uses time and space to do so. But computational complexity is not interesting only for its ``positive'' and ``plain'' side: the most beautiful results in complexity are those which show the equivalence of seemingly different resources, and, of course, negative results which show intrinsic limitations on what can be solved efficiently. It is our aim to recast those results to talk about all systems capable of solving problems, i.e. machines and minds.

\section{The problem of huge lookups}

Back to our notion of solvability. We have mentioned that solving a problem by accident is not a solution to the problem. However, we would also like to pinpoint that not every solution should be a good one, in the context of philosophy of mind. Of course, if one were to find a solution to an $\NP$-complete problem, that solution could be applied to all the problems in the same class. However, a brute-force solution is still a solution, albeit not really a good one. One of the issues in computational complexity is to find a polynomial-time algorithm for such cases. We would expect such a solution to be an elegant, applicable one.

That's where the concept of a huge lookup table comes into play. A \textit{lookup table} is simply a stored table of values, but it can take other forms as well. One of its most important alternative forms is a huge list of IF-THEN-ELSE statements covering all possible solutions. If one had a piece of code that would effectively be just a huge lookup table, it would feel like cheating, for it packed all of the actual computation in its code, so it is \textit{de facto} just routing, and thus achieving polynomial time. 

Nice examples of lookup tables are multiplication tables kids use to learn how to multiply. They memorize it by heart to jump-start a recursive procedure. Give a kid to calculate 771 times 81. The kid will probably do something like: ``I multiply 771 by 1, which is just 771, and then I multiply 771 by 8 and I get 6168, put a zero at the end of it to get 61680 and add this to 771 to get the result 62451.'' Notice the number of steps here, and the (perfectly valid) shortcuts the kid will take to save time. At the end (the shift and the addition), there is recursion, but at the beginning, not so much: the kid multiplied 771 by 8, and she did it by knowing by heart what 7 times 8 is.  By knowing this value by heart, she was effectively using a lookup table and speeding up her computation. If she had not just possessed a lookup table of single-digit multiplication, but of, say, all two and three-digit multiplications, she would not need to solve the problem at all: she would just look up the solution in the big lookup table in her mind. The time it would take to solve it would be instantaneous, i.e. constant O($1$). 

So are lookup tables a way to bypass complexity? No. A lookup table just shows that by having enough computational space (memory), you can simply reduce the computing or inferring to a minimum. What it says is that time is subsumed by space. 

The philosophical intuition here is that even though it computes, it does not really \textit{solve} the problem, the same way one’s spouse dying is not really a right solution to one’s former marital problems. We would like to strengthen this claim further: even though the code itself could work in polynomial time, if it wasn't elegant, efficient, and self-sufficient, it would most definitely \textit{compute}, but it would not really \textit{solve} a problem. Imagine an algorithm that solves an $\NP$-complete problem in polynomial time. Both of your authors have worked as programmers, and if there was one thing we learned, is that code lines themselves do not matter, the only thing that matters is the execution time. However, as philosophers, we think differently: a piece of code that mostly \textit{computes}, rather than \textit{solves}, is not relevant for the philosophy of mind, and it might even turn out to be less meaningful for computational complexity as well.

\textit{Turing test} (originally called by Turing \cite{turing} an \textit{imitation game}) is a standard well-known test of a machine's ability to exhibit intelligent behavior, that would be indistinguishable from that of a human being. We are not for the moment interested in the fact whether we would be right to ascribe consciousness, qualia, or intelligence to such a machine, but in the practical issue: could a program that passed (a strong version of) the Turing test \textit{actually} be written \cite{aaronson2011}? However, the more important question here is what does it actually mean that there is a task that humans can perform but computers cannot. Aaronson \cite{aaronson2011} states that in practice, people judge each other to be conscious after interacting for a very short time. A human being can very soon figure out that he or she is talking to a chatbot on a certain website or in an application. That means there is a finite upper bound on the number of bits (for example, $10^{20}$ bits) of information that two people – A and B – would realistically exchange, for B to conclude that A is conscious. If we were to imagine a lookup table that would hold all the history H of A and B's conversation and the action $f_B(H)$ that $B$ would take next given that history, such a table would be, of course, mostly filled with meaningless nonsense. If $A$ states that it is sunny outside, B might continue the conversation in a natural way, but $B$ might also dance hula, eat candy using his/her toes, or ignore the question and read Wittgenstein. Such a table would be too large to fit inside the observable universe, but Aaronson \cite{aaronson2011} points out that it would still be \textit{finite}, since we have presupposed an upper bound on the conversation length, and that function $f_B$ is \textit{computable}. If one wants to consider the brain an analog to an efficient Turing machine, and if we presuppose that the mentioned astronomical lookup table is essentially the best one could do in order to simulate the human brain, we still -- to our current mathematical knowledge -- cannot prove that such a claim.

This astronomical lookup table immediately evokes Searle's \textit{Chinese Room argument} \cite{searle}, whose purpose was to show that a computer manipulating symbols cannot be presumed conscious since there is no real understanding of the process, i.e. \textit{intentionality}. Aaronson \cite{aaronson2011} postulates that a giant lookup table would immediately seem irrelevant to intelligence. It is just a huge collection of symbols and nonsense and not a real consciousness. However, if one imagines a \textit{compact, efficient} computer program passing the Turing test, one would have to explain such efficiency by positing that the program includes capacities for learning and reasoning, abstract representations, and other mental phenomena we usually attribute to the human mind. That would, Aaronson concludes, seem to imply that there is a \textit{metaphysical} significance between \textit{polynomial} and \textit{exponential} complexity. But this is, we would like to point out, a consequence of such a huge collection only \textit{computing}, rather than \textit{solving} the Chinese Room problem.

Parrberry \cite{parberry} has pinpointed a flaw in Searle's \cite{searle} reasoning that we would like to once again emphasize: \textit{there is no such imaginary (possible) world where computation costs nothing}.  The Chinese Room is actually an impossible scenario. For the sake of the argument, let's call the person inside the room \texttt{Searle}. Since \texttt{Searle}'s lookup table grows exponentially in the size of the queries, it cannot be easily manipulated by the \texttt{Searle}'s polynomial mind. By polynomial mind, we actually understand the limitations of \texttt{Searle}'s mental capabilities of processing and feasibly manipulating the exponential lookup table, in a way that the person submitting the query would be satisfied by how fast she gets the answer. This is important since the whole flavor of the Chinese Room argument can be understood as variation of the Turing test. If the processing speed of the room is either too quick or too slow, a confusion about the person inside the room speaking or not speaking Chinese cannot be claimed, rendering the whole argument pointless.

\section{Computationalism or complexity}

One of the most popular theories in philosophy of mind is computationalism, which is also especially endearing to computer scientists. The CTM or the \textit{computational theory of mind} presupposes that the mind itself is a computational system. CTM argues that the mind computes in the relevant sense, and relates such a computational description with neurophysiology. Warren McCulloch, a neurophysiologist and a cyberneticist, and Walter Pitts, a logician first suggested that neural activity was computational, thus establishing the notion of neural networks as a computational model. McCulloch and Pitts \cite{mcculloch-pitts} argued that neural computation could explain cognition, which was further developed in philosophy, cognitive science, and computer science. However, if one looks at relevant and/or popular overviews and historical accounts of computationalism, there is practically no mention of computational complexity whatsoever. For example, SEP \cite{sep} does not mention it at all,\footnote{There is a separate page on computational complexity, but it seems that it’s an isolated case.} while IEP \cite{iep} mentions only that the universal Turing machine was never used directly to write computational models of cognitive tasks, and its role may be seen as \textit{merely instrumental} (emphasis added) in analyzing the computational complexity of algorithms posited to explain these tasks. We would like to state that computational complexity is \textit{essentially} instrumental, rather than \textit{merely} instrumental, since any real possibility of developing a sentient or conscious mind would depend on computational complexity having polynomial complexity.

Little is known about consciousness, and therefore philosophers love mental experiments that are showing idealized cases that cannot be proven wrong. We do not care whether the mind is really a computational artifact, but there is one important thing regarding the human brain, which goes against the grain of virtually all CTMs: it is \textit{finite}. Being a finite object, and since consciousness is directly connected to the brain, we can safely deduce that consciousness is finite. It does not matter now that the computer is a finite object as well, such equalities do not interest us here.\footnote{We would like to pinpoint though that there is an ontological equality of a brain as a computation device and a computer if one looks at the cybernetic notion of a same process. McCulloch \cite{mcculloch-pitts} as a cyberneticist wanted to emulate a neurological process using a machine, acting under the heritage of cybernetics, which aimed at establishing process equalities between various biological and non-biological systems.} One of the biggest former examples of ignoring computational complexity comes from Penrose \cite{penrose}, even though he was attacking CTM. Penrose asserts that consciousness is connected to non-computable processes that cannot be simulated by a Turing machine. Aaronson \cite{aaronson2013} answers that if he really wanted to speculate about the impossibility of simulating the brain on a computer, he should talk about \textit{complexity} rather than \textit{computability}. The mentioned universe-sized lookup table could always simulate a conscious being. Albeit universe-sized, such a table would still be \textit{finite}.

There are two important questions here. First, it seems intuitive that we would ascribe consciousness to a computation that is fast and efficient. It does not have to be a literal computation in the common sense of the word, it can be an answer to a simple chat question. But there is a \textit{finite} number of bits interchanged humans need to presume that the other side possesses consciousness. We have mentioned that it seems that time is more important in such a decision than the notion of memory, i.e. space. So, polynomial time is not only a different ontology of time, compared to, say, exponential complexity, but also more important in the philosophical sense of deciding whether a system is conscious or not. So, not only do we have computational space and time hierarchies, but also hierarchies of time that are more important than those of space as a necessary prerequisite for deciding about a person's or computer's intelligence. Philosophers love zombies. A zombie that would take ages to respond would behave like a slow computer, and it would be difficult to ascribe any sentience. However, a fast and efficient talkative zombie, that's quite a mental experiment.

Second, there could be $\NP$-complete problems that would emulate human creativity. We could be able to generate perfect poems if creativity or aesthetics could be (quickly) computable (and if $\P = \NP$). However, even if they do not exist, this tells nothing about human intelligence. In \textit{I, Robot} (2004), detective Spooner asks a robot: ``Can a robot write a symphony? Can a robot turn a canvas into a beautiful masterpiece?'', to which the robot answers: ``Can you?''. Human creativity is not a criterion for intelligence, at least not a necessary one, but it might be a sufficient one. And, if $\P = \NP$, and humans are still not good at solving $\NP$-complete problems, machines could be better than humans in such cases. What would be our epistemological stance towards deciding what constitutes \textit{real} knowledge and understanding? Certainly not aesthetics, but it seems that difficult computations might be irrelevant as well.

If we are building systems that emulate the human mind, and if we claim that the human mind is actually a computer, the question of artificial intelligence would not be a question of solving knapsack or traveling salesman problems. It would (probably) be a matter of having a conversation, which brings us back to Turing’s Imitation game \cite{turing}. Therefore, not only that computational complexity can provide us with different temporal ontologies, but it seems that different computational problems carry little or much weight for deciding whether a system is intelligent or not. A strong $\NP$-complete problem could be used for any similar problems, so a knapsack solution could be used to write a poem or carry a philosophical conversation. But solving an $\NP$-hard problem such as a non-decision traveling salesman does not seem relevant. We might posit another claim here: $\NP$-complete problems are more important in philosophy of mind and artificial intelligence because their solution could be used for different sorts of problems, but specific $\NP$-hard problems seem irrelevant to deciding whether a system is intelligent or not, or even a human being. Therefore, even if we have polynomial solutions to both an $\NP$-hard and an $\NP$-complete problem, an $\NP$-complete solution is what interests us more, and seems to carry a different ontological status as well. Furthermore, $\NP$-complete problems are important for artificial intelligence \textit{because} they are important for the philosophy of mind, in a sense that anything mental pertaining to complexity will inherently need to be addressed when rebuilding the mind in the context of artificial intelligence.

So why bother with $\P$ vs. $\NP$ in the first place? First, the notion of computational complexity is ignored in philosophical writings and thought experiments. A thought experiment rarely takes into account whether something is really usable in the real world. I do not care that you can provide me with a solution to my housing problem in a couple of millennia, and such a scenario tells nothing about the metaphysical difference between a vast, exponential thought system and an efficient, polynomial one. We would like to pinpoint that such an ontological difference should be taken into account when talking about brain simulations or the possibility of (hard) artificial intelligence.

\section{Conclusion}

Even though the main goal of this paper is to show the importance of computational complexity not only in philosophy of mind but in philosophy in general, an important side-result is to show how the notion of \textit{computability} bears little to no consequences compared to the revised Church-Turing thesis comprising \textit{solvability}. Multiple theories of mind are comparing the human mind to a computer and positing that mental operations are computations. However, we believe that the main explanatory gap lies in the fact that something being just a computation is not enough for intelligence. The standard definition of an explanatory gap \cite{levine} in philosophy of mind is the difficulty in explaining how physical properties give rise to subjective conscious experience. We would like to strengthen this claim further: not only there is a gap between physical properties and qualia, i.e. how things feel when experienced, but between various physical ways of producing such phenomena. However, to our luck, such a gap in the former case can be explained by differentiating between computability and solvability. We would, it seems, intuitively ascribe consciousness and thus qualia to a system that solves problems, rather than a system that just computes. 

Let us look more closely at the mind and computer similarities. First, we need to pinpoint that inefficient and slow computations do not automatically imply non-consciousness and vice versa. Ask an Italian how to make carbonara, you'll get your answer in an instant, but not as fast if you ask an Indian. The person might even stall, but you cannot really reset mental processes as easily as you can in Unix machines. And, of course, a fast and efficient computation does not imply sentience as well since my computer can multiply 10-digit numbers with ease. So, fast and efficient computation -- or as we have called it -- polynomial ontology -- is not a sufficient condition for consciousness, but it seems to be a necessary one for selected problems.

Computational complexity is important because there seems to be an ontological and epistemological difference in various ways time and space can be consumed. Both computational time and space lie at the foundations of both the notion of computability and the notion of solvability. The main difference is that we cannot exactly talk about minds without taking into account computational complexity, and, we would like to add, it is \textit{impossible} to talk about intelligence without taking into account the next step in the computational hierarchy: solvability.

\bibliography{bibl.bib}
\bibliographystyle{plain}

\end{document}